\documentclass[aps,prb,twocolumn,groupedaddress,showpacs]{revtex4}
\usepackage{graphicx}
\usepackage{dcolumn}
\usepackage{bm}

\begin{document}

\title{%
Deformation of SU(4) singlet spin-orbital state due to Hund's rule coupling}

\author{Hiroaki Onishi}

\affiliation{%
Advanced Science Research Center,
Japan Atomic Energy Agency,
Tokai, Ibaraki 319-1195, Japan}

\date{January 5, 2007}

\begin{abstract}
We investigate the ground-state property and the excitation gap of
a one-dimensional spin-orbital model
with isotropic spin and anisotropic orbital exchange interactions,
which represents the strong-coupling limit of
a two-orbital Hubbard model including the Hund's rule coupling ($J$)
at quarter filling,
by using a density-matrix renormalization group method.
At $J$=0, spin and orbital correlations coincide with each other
with a peak at $q$=$\pi/2$,
corresponding to the SU(4) singlet state.
On the other hand, spin and orbital states change in a different way
due to the Hund's rule coupling.
With increasing $J$,
the peak position of orbital correlation changes to $q$=$\pi$,
while that of spin correlation remains at $q$=$\pi/2$.
In addition, orbital dimer correlation becomes robust
in comparison with spin dimer correlation,
suggesting that quantum orbital fluctuation is enhanced
by the Hund's rule coupling.
Accordingly, a relatively large orbital gap opens
in comparison with a spin gap,
and the system is described by an effective spin system
on the background of the orbital dimer state.
\end{abstract}



\maketitle

\section{Introduction}

It has been recognized that
orbital degree of freedom plays a crucial role
in the emergence of exotic magnetism
in strongly correlated electron systems.
The interplay of spin and orbital degrees of freedom
triggers off a variety of orbital ordering that stabilizes
a certain spin structure on the orbital-ordered background,
leading to complex ordered phases,
as frequently observed in transition metal oxides
and $f$-electron compounds.~\cite{Imada1998,Hotta2006}

In addition to the appearance of diverse spin-orbital ordered states,
the combined quantum effects of spin and orbital degrees of freedom
yields a possibility of novel quantum critical
and quantum disordered states.~\cite{Feiner1997}
In this context,
a high symmetric SU(4) exchange model,
which describes a spin-1/2 system
coupled with a pseudospin-1/2 for two-fold orbital degeneracy,
has been one of the intriguing subjects from a theoretical
viewpoint.~\cite{Sutherland1975,Affleck1986,
Yamashita1998,Frischmuth1999,Li1998}
In such a high symmetric case,
spin, orbital, and combined spin-orbital degrees of freedom
play an identical role.~\cite{Li1998}
In particular, the one-dimensional model is
Bethe ansatz solvable,~\cite{Sutherland1975}
and the quantum critical behavior of an SU(4) singlet state
has been revealed by analytical and numerical
investigations.~\cite{Affleck1986,Yamashita1998,Frischmuth1999}
In fact, correlation functions show power-law decay
with a critical exponent 3/2,
and the elementary excitation is gapless.

The SU(4) spin-orbital exchange model represents
the strong-coupling limit of a two-orbital Hubbard model
at quarter filling,
in which electrons hop between the same types of orbitals
with equal amplitude and the Hund's rule coupling is ignored.
Indeed, such a simplified model is an appropriate starting point
to gain deep insight into the complex quantum fluctuations
of spin and orbital degrees of freedom.
However, in actual materials,
the highest SU(4) symmetry is likely to be broken down,
thus it is quite important to clarify the effects of
possible symmetry breaking.
For instance,
intensive studies have been devoted to a one-dimensional
SU(2)$_{\rm spin}$$\times$SU(2)$_{\rm orbital}$ exchange model
and the ground-state phase diagram has been
revealed.~\cite{Kolezhuk1998,Pati1998,Azaria1999,
Tsukamoto2000,Yamashita2000,Itoi2000,Azaria2000}
Around the SU(4) point,
a quantum critical phase extends in one side
next to the SU(4) point,~\cite{Yamashita2000,Itoi2000,Azaria2000}
while in the other side there occurs a gapped phase,
where spin- and orbital-singlet dimers are formed
in an alternating pattern.~\cite{Kolezhuk1998}

In a more realistic situation, however,
the Hund's rule coupling causes
anisotropic exchange interactions in the orbital part,
and the SU(4) symmetry is broken down to
SU(2)$_{\rm spin}$$\times$U(1)$_{\rm orbital}$.~\cite{Arovas1995,
Yamashita1998,Lee2004,Xavier2006,Onishi2006}
To clarify the effects of such symmetry breaking,
an SU(4) Hubbard model perturbed by the Hund's rule coupling
has been analyzed by means of renormalization-group and bosonization
methods.~\cite{Lee2004}
It has been proposed that the excitation gap opens
for an arbitrarily small Hund's rule coupling,
while the opening excitation gap is exponentially small
as a function of the Hund's rule coupling,
corresponding to a generalized type of
Kosterlitz-Thouless transition.~\cite{Itoi1999}

On the other hand,
it has been common understanding that the Hund's rule coupling
in multi-orbital systems plays an important role in itinerant
ferromagnetism.~\cite{Slater1936,VanVleck1953}
In fact, it has been shown that
the Hund's rule coupling induces a ground-state transition
from a paramagnetic (PM) state to a ferromagnetic (FM) state
in the strong-coupling region,~\cite{Gill1987,Kuei1997,Sakamoto2002}
but there, the discussion has been focused on the appearance of
ferromagnetism itself and the property of spin and orbital states
in the PM phase has not been clear yet.
Thus, we believe that it is an intriguing issue to clarify
how the SU(4) singlet state is deformed to
a spin- and orbital-singlet state due to the Hund's rule coupling.

In this paper,
we investigate the ground-state property and the excitation gap
of a one-dimensional spin-orbital model
with the SU(2)$_{\rm spin}$$\times$U(1)$_{\rm orbital}$ symmetry,
by using numerical techniques.
When the Hund's rule coupling is zero,
the system has the SU(4) symmetry,
and spin and orbital correlations have a peak at $q$=$\pi/2$.
On the other hand,
the Hund's rule coupling induces
FM and antiferro-orbital (AFO) interactions,
leading to the characteristic change of the peak position
of orbital correlation to $q$=$\pi$,
while that of spin correlation remains at $q$=$\pi/2$ in the PM phase.
Moreover, we observe the stabilization of an orbital dimer state
with a significant orbital excitation gap.
Taking into account the robust orbital dimerization,
the low-energy physics is described by an effective spin system
on the background of the orbital dimer state.

The organization of this paper is as follows.
In Sec.~II,
we describe the model Hamiltonian and the numerical method.
Starting from a two-orbital Hubbard model,
a spin-orbital exchange model in the strong-coupling limit is introduced.
In Sec.~III,
we show our numerical results of physical quantities such as
correlation functions and excitation gaps.
We discuss distinctive changes of spin and orbital states
from the SU(4) singlet state due to the Hund's rule coupling.
Finally, we summarize the paper in Sec.~IV.

\section{Model and Method}

Let us consider doubly degenerate orbitals
on a one-dimensional chain with $N$ sites
including one electron per site (quarter filling).
The two-orbital Hubbard model is described by
\begin{eqnarray}
\label{eq:H-Hubbard}
 \tilde{H}
 &=&
 t\sum_{i,\tau,\sigma}
 (d_{i\tau\sigma}^{\dag} d_{i+1\tau\sigma}+\mbox{h.c.})
 + U \sum_{i,\tau} \rho_{i\tau\uparrow} \rho_{i\tau\downarrow}
\nonumber\\
 &&
 + U'\sum_{i,\sigma,\sigma'} \rho_{i\alpha\sigma} \rho_{i\beta\sigma'}
 + J \sum_{i,\sigma,\sigma'}
 d_{i\alpha\sigma}^{\dag} d_{i\beta\sigma'}^{\dag}
 d_{i\alpha\sigma'} d_{i\beta\sigma}
\nonumber\\
 &&
 + J'\sum_{i,\tau \ne \tau'} 
 d_{i\tau\uparrow}^{\dag} d_{i\tau\downarrow}^{\dag}
 d_{i\tau'\downarrow} d_{i\tau'\uparrow},
\end{eqnarray}
where $d_{i\tau\sigma}$ ($d_{i\tau\sigma}^{\dag}$)
is an annihilation (creation) operator for an electron
with spin $\sigma$(=$\uparrow,\downarrow$)
in orbital $\tau$(=$\alpha,\beta$) at site $i$,
$\rho_{i\tau\sigma}$=$d_{i\tau\sigma}^{\dag}d_{i\tau\sigma}$,
and $t$ is the hopping amplitude.
Here, we assume that
electrons hop between the same types of orbitals with equal amplitude.
Note that, in general, the hopping amplitudes of multi-orbital systems
are evaluated from the overlap integral
between orbitals.~\cite{Slater1954,Dagotto2001}
The present simple form of the hopping amplitudes
represents the system with, for instance,
($p_x,p_y$) orbitals or ($d_{yz},d_{zx}$) orbitals
on a chain along the $z$ axis.

In the interaction terms,
$U$, $U'$, $J$, and $J'$ denote
intra-orbital Coulomb,
inter-orbital Coulomb,
inter-orbital exchange (Hund's rule coupling),
and inter-orbital pair hopping interactions, respectively.
These Coulomb integrals are all positive.
Note that the spin diagonal part of the Hund's rule coupling
causes an attractive interaction,
so that the inter-orbital Coulomb interaction is effectively
reduced to $U'$$-$$J$.
To ensure that the total inter-orbital Coulomb interaction
is repulsive, we consider the region of $J$$<$$U'$.
We also note that the relation $U$=$U'$+$J$+$J'$ holds
due to the rotational invariance in the local orbital space,
and $J$=$J'$ due to the reality of the orbital
wavefunction.~\cite{Dagotto2001}
When the Hund's rule coupling is zero,
i.e., $U$=$U'$ and $J$=$J'$=0,
the system possesses the SU(4) symmetry.~\cite{Lee2004}

In order to consider an effective model
in the strong-coupling limit,
as usual, we start from the ground state of the atomic limit $t$=0,
and treat the electron hopping
as a perturbation.~\cite{Kugel1972,Kugel1973}
In the ground state of the atomic limit,
each site is occupied by one electron,
leading to four-fold degenerate states, as shown in Fig.~1(a).
Then, the electron hopping causes virtual two-electron states,
as shown in Fig.~1(b).
There appear four eigenenergy states,
as listed in Table~I.
Among them, the lowest energy state is spin triplet,
and has energy $U'$$-$$J$.
Thus, the Hund's rule coupling stabilizes spin polarized states.
The other three states are spin singlet.
Note that the second- and third-lowest energy states
are degenerate because of the relation $U$=$U'$+$J$+$J'$.
Note also that all these four levels are degenerate
when $U$=$U'$ and $J$=$J'$=0.

\begin{figure}[t]
\includegraphics[width=\linewidth]{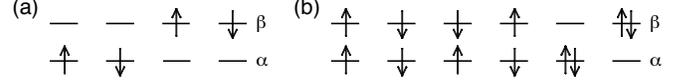}
\caption{
Spin-orbital configuration in (a) one-electron state
and (b) two-electron state.
}
\end{figure}

\begin{table}[t]
\caption{
Eigenenergy and eigenstate of two-electron state
in the ascending order of eigenenergy,
where $\vert \tau\sigma \rangle$ denotes
the spin $\sigma$ and orbital $\tau$ state.
}
\begin{ruledtabular}
\begin{tabular}{cc}
eigenenergy & eigenstate \\
\hline
$U'$$-$$J$ &
$\vert \alpha\uparrow \rangle \vert \beta\uparrow \rangle$,
$\vert \alpha\downarrow \rangle \vert \beta\downarrow \rangle$,
$\frac{1}{\sqrt{2}}
 (\vert \alpha\uparrow \rangle \vert \beta\downarrow \rangle$%
 +$\vert \alpha\downarrow \rangle \vert \beta\uparrow \rangle)$ \\
$U'$$+$$J$ &
$\frac{1}{\sqrt{2}}
 (\vert \alpha\uparrow \rangle \vert \beta\downarrow \rangle$%
 $-$$\vert \alpha\downarrow \rangle \vert \beta\uparrow \rangle)$ \\
$U$$-$$J'$ &
$\frac{1}{\sqrt{2}}
 (\vert \alpha\uparrow \rangle \vert \alpha\downarrow \rangle$%
 $-$$\vert \beta\uparrow \rangle \vert \beta\downarrow \rangle)$ \\
$U$$+$$J'$ &
$\frac{1}{\sqrt{2}}
 (\vert \alpha\uparrow \rangle \vert \alpha\downarrow \rangle$%
 +$\vert \beta\uparrow \rangle \vert \beta\downarrow \rangle)$ \\
\end{tabular}
\end{ruledtabular}
\end{table}

Now we apply the degenerate perturbation theory
by considering the second-order processes of the electron hopping,
and obtain an effective spin-orbital exchange model, such as
\begin{eqnarray}
\label{eq:H-so-original}
 H&=&\sum_{i} H_{i,i+1},
\nonumber\\
 H_{i,j}&=&
 A
 \left(
   {\bf S}_{i}\cdot{\bf S}_{j}+\frac{3}{4}
 \right)
 \left(
   {\bf T}_{i}\cdot{\bf T}_{j}-\frac{1}{4}
 \right)
\nonumber\\
 &&
 +B
 \left(
   {\bf S}_{i}\cdot{\bf S}_{j}-\frac{1}{4}
 \right)
 \left(
   {\bf T}_{i}\cdot{\bf T}_{j}
   -2T_i^z T_j^z + \frac{1}{4}
 \right)
\nonumber\\
 &&
 +C
 \left(
   {\bf S}_{i}\cdot{\bf S}_{j}-\frac{1}{4}
 \right)
 \left(
   {\bf T}_{i}\cdot{\bf T}_{j}
   -2T_i^x T_j^x + \frac{1}{4}
 \right)
\nonumber\\
 &&
 +D
 \left(
   {\bf S}_{i}\cdot{\bf S}_{j}-\frac{1}{4}
 \right)
 \left(
   {\bf T}_{i}\cdot{\bf T}_{j}
   -2T_i^y T_j^y + \frac{1}{4}
 \right),
\end{eqnarray}
where
\begin{equation}
 {\bf S}_{i}=
 (1/2)\sum_{\tau,\sigma,\sigma'}
 d_{i\tau\sigma}^{\dag}
 \mbox{\boldmath $\sigma$}_{\sigma\sigma'}
 d_{i\tau\sigma'}
\end{equation}
is the $S$=1/2 spin operator and
\begin{equation}
 {\bf T}_{i}=
 (1/2)\sum_{\tau,\tau',\sigma}
 d_{i\tau\sigma}^{\dag}
 \mbox{\boldmath $\sigma$}_{\tau\tau'}
 d_{i\tau'\sigma}
\end{equation}
indicates the $T$=1/2 pseudospin operator representing two orbitals,
with the Pauli matrix $\bm{\sigma}$.
Concerning the exchange interactions $A$$\sim$$D$
in Eq.~(\ref{eq:H-so-original}),
each of them arises from the corresponding eigenenergy of
the virtual two-electron state,
and they are expressed as
\begin{eqnarray}
 A &=& 4t^2/(U'-J),
\\
 B &=& 4t^2/(U'+J),
\\
 C &=& 4t^2/(U-J') = B,
\\
 D &=& 4t^2/(U+J') = 4t^2/(U'+3J).
\end{eqnarray}
Note that the relation $A$$>$$B$$>$$D$$>$0 holds
in the realistic parameter region of 0$\leq$$J$$<$$U'$.
Hereafter, we take $t^2/U'$ as the energy unit.
Normalized by $t^2/U'$,
the exchange interactions are parametrized by $J/U'$.
Since we focus on the region of 0$\leq$$J$$<$$U'$,
we set $U'$=1 for simplicity and consider the variation of $J$
within 0$\leq$$J$$<$1.

Here, let us discuss the symmetry of the spin-orbital model.
When $U$=$U'$ and $J$=$J'$=0,
the spin-orbital model is reduced to
\begin{equation}
 H_{i,j}^{(0)}=
 \frac{8t^2}{U}
 \left({\bf S}_i \cdot {\bf S}_j +\frac{1}{4}\right)
 \left({\bf T}_i \cdot {\bf T}_j +\frac{1}{4}\right),
\end{equation}
which has the SU(4) symmetry.~\cite{Yamashita1998,Li1998}
On the other hand,
in the spin-orbital model of Eq.~(\ref{eq:H-so-original}),
we can clearly see that
exchange interactions of the spin part are isotropic,
while those of the orbital part show anisotropic forms
according to the energy level of the virtual two-electron state.
Note that the second and third terms have
the same energy denominator due to the relation $U$=$U'$+$J$+$J'$,
so that exchange interactions in the orbital part
become isotropic in the $zx$ plane.
In fact, $T_{\rm tot}^{y}$=$\sum_{i}T_{i}^{y}$ commutes with $H$,
and becomes a good quantum number.
Concerning spin degree of freedom,
total spin $S_{\rm tot}$ and $S_{\rm tot}^{z}$=$\sum_{i}S_{i}^{z}$
are good quantum numbers.
Thus, the SU(4) symmetry is broken down to
SU(2)$_{\rm spin}$$\times$U(1)$_{\rm orbital}$
due to the Hund's rule coupling.

\begin{figure}[t]
\includegraphics[width=\linewidth]{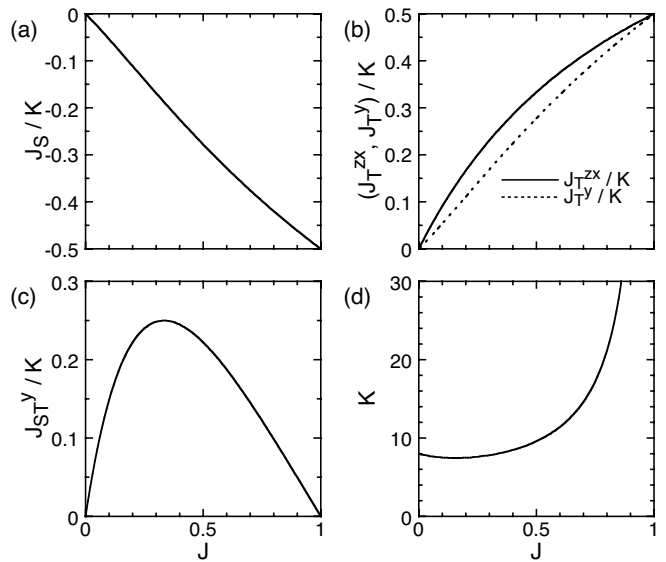}
\caption{
Exchange interactions in the spin-orbital model (\ref{eq:H-so}):
(a) $J_{S}/K$,
(b) $J_{T}^{zx}/K$, $J_{T}^{y}/K$,
(c) $J_{ST}^{y}/K$,
and (d) $K$,
as a function of $J$.
}
\end{figure}

In order to discuss the change of the spin-orbital state
from the SU(4) symmetric case,
it is useful to rewrite the Hamiltonian
by the sum of the SU(4) symmetric part and the other parts,
such as
\begin{eqnarray}
\label{eq:H-so}
 H_{i,j}
 &=&
 K ({\bf S}_i \cdot {\bf S}_j +1/4)({\bf T}_i \cdot {\bf T}_j +1/4)
\nonumber\\
 &&
 +J_{S} {\bf S}_i \cdot {\bf S}_j
 +J_{T}^{zx} (T_i^x T_j^x + T_i^z T_j^z)
 +J_{T}^{y} T_i^y T_j^y
\nonumber\\
 &&
 +J_{ST}^{y} {\bf S}_i \cdot {\bf S}_j T_i^y T_j^y,
\end{eqnarray}
where the exchange interactions are given by
\begin{eqnarray}
 K &=& A+D,
\\
 J_{S} &=& (-A+B)/2,
\\
 J_{T}^{zx} &=& (A-D)/2,
\\
 J_{T}^{y} &=& (A-B)/2,
\\
 J_{ST}^{y} &=& 2(B-D).
\end{eqnarray}
We can easily see that
$J_{S}$ is negative,
while $J_{T}^{zx}$ and $J_{T}^{y}$ are positive,
indicating that FM and AFO interactions are induced by
the Hund's rule coupling.
We note the relation of
\begin{equation}
\label{eq:relation}
 J_{T}^{y} + J_{ST}^{y}/4 = J_{T}^{zx},
\end{equation}
among anisotropic orbital exchanges.

In Figs.~2(a)-(d),
exchange interactions are plotted as a function of $J$.
At $J$=0,
exchange interactions other than $K$ are equal to zero,
and the system possesses the SU(4) symmetry.
On the other hand,
the Hund's rule coupling induces
FM and AFO interactions,
which should cause the deviation of the spin-orbital state
from the SU(4) symmetric case,
as will be discussed in detail in the next section.

We analyze the spin-orbital model (\ref{eq:H-so}),
or equivalently (\ref{eq:H-so-original}),
by a density-matrix renormalization group (DMRG)
method.~\cite{White1992,Schollwock2005}
The finite-system algorithm is employed
with open boundary conditions.
We keep up to 800 states in the renormalization step
and the truncation error is kept around 5$\times$$10^{-6}$ or smaller.
We also use a Lanczos method
for the analysis of a four-site periodic chain.

\section{Results}

\subsection{Four-site chain}

\begin{figure}[t]
\includegraphics[width=\linewidth]{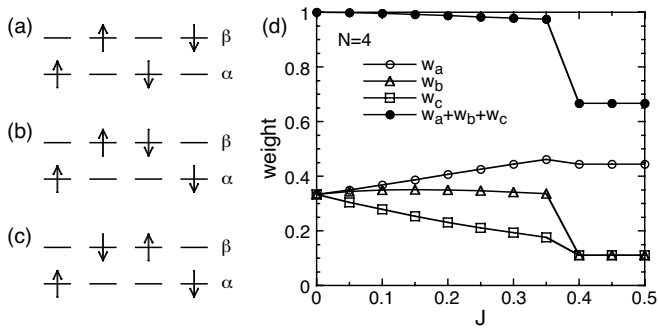}
\caption{
(a)-(c) Three classes of relevant spin-orbital configurations
in the ground state of the four-site system.
(d) Weight of each class of Figs.~3(a)-(c) in the ground state
as a function of $J$.
}
\end{figure}

To grasp the fundamental property of the spin-orbital configuration
in the ground state,
first we consider a four-site periodic chain,
which is a minimal model to form a unique SU(4) singlet ground state
at $J$=0.~\cite{Li1998}
The SU(4) singlet state for $N$=4 is composed of
different four spin-orbital configurations at four sites,
expressed as
\begin{equation}
 \vert {\rm SU(4)} \rangle =
 (1/\sqrt{24}) \sum_{i \neq j \neq k \neq l}
 d_{i\alpha\uparrow}^{\dag}d_{j\alpha\downarrow}^{\dag}
 d_{k\beta\uparrow}^{\dag}d_{l\beta\downarrow}^{\dag}
 \vert 0 \rangle,
\end{equation}
where $\vert 0 \rangle$ is the vacuum state and
the summation is taken over all permutations
in terms of the site index.
Note that $\vert {\rm SU(4)} \rangle$
consists of 24 states with the same weight.
For finite $J$, however, the ground state is not represented by
$\vert {\rm SU(4)} \rangle$ itself.
In fact, the ground state is changed to a spin- and orbital-singlet state,
which is well described by the superposition
of the 24 states included in $\vert {\rm SU(4)} \rangle$,
but the 24 states are split into three classes according to the weight
in the ground state,~\cite{Xavier2006,Onishi2006}
as shown in Figs.~3(a)-(c).
Note that each class has eight equivalent states
due to the translation and the reversal of orbitals.
In Fig.~3(d),
we show the $J$ dependence of the weight of each class $m$
in the ground state, defined by
\begin{equation}
 w_m=\sum_{i\in m}
 \vert \langle \phi_{i} \vert \psi_{\rm G} \rangle \vert^2,
\end{equation}
where $\psi_{\rm G}$ is the ground state and
$\phi_{i}$ denotes the basis.
At $J$=0,
each class contributes to the ground state with the equal weight
due to the SU(4) singlet ground state.
On the other hand,
with increasing $J$,
$w_{\rm a}$ gradually increases,
while $w_{\rm b}$ does not change so much and $w_{\rm c}$ decreases.
Note that the total weight of the three classes
remains almost unity even when $J$ is increased,
but it abruptly decreases at around $J$=0.4,
since the ground state turns
from the singlet state into a FM state.

Here, it is worth noting that
each of the three classes of Figs.~3(a)-(c) is identified by
a distinct peak position in spin and orbital structure factors,
denoted by $(q_{\rm s},q_{\rm o})$:
$(\pi/2,\pi)$ for the class (a),
$(\pi/2,\pi/2)$ for the class (b), and
$(\pi,\pi/2)$ for the class (c).
In the class (a),
a parallel spin arrangement is stabilized by
the FM interaction due to the Hund's rule coupling,
while the pairs of parallel spins point opposite directions
to form a spin-singlet ground state for small $J$.
Concerning the orbital state,
an AFO configuration appears so as to avoid the energy loss
due to the double occupancy in the virtual hopping process.
On the other hand, the class (b) also exhibits
a parallel spin arrangement,
but there occurs a partly ferro-orbital configuration,
leading to relatively small $w_{\rm b}$ comparing with $w_{\rm a}$.
The class (c) is more unfavorable
due to a fully antiferromagnetic configuration.

\subsection{DMRG results}

\begin{figure}[t]
\includegraphics[width=\linewidth]{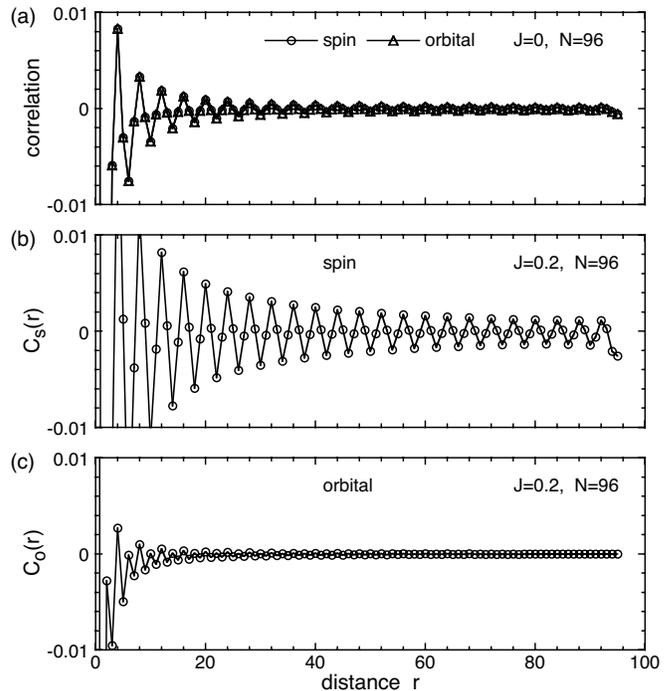}
\caption{
(a) Spin and orbital correlation functions at $J$=0.
(b) Spin and (c) orbital correlation functions at $J$=0.2.
}
\end{figure}

From the discussion within the small four-site chain
in the previous subsection,
we can intuitively understand the effect of the Hund's rule coupling
to cause the characteristic change of the spin-orbital state,
but it is highly required to clarify what types of correlations
develop in the thermodynamic limit.
For this purpose,
here we move on to our DMRG results for longer chains.
In Fig.~4, we show spin and orbital correlation functions,
defined by
\begin{eqnarray}
 C_{\rm s}(r)
 &=&
 (1/N_r)\sum_{|i-j|=r}
 \langle S_i^z S_j^z \rangle,
\\
 C_{\rm o}(r)
 &=&
 (1/N_r)\sum_{|i-j|=r}
 \langle T_i^z T_j^z \rangle,
\end{eqnarray}
where $N_r$ is the number of site pairs $(i,j)$ with $r$=$|i$$-$$j|$,
and $\langle\cdots\rangle$ denotes the expectation value.
Note that we average over all pairs of sites separated by distance $r$
in order to reduce the effect of open boundaries.
At $J$=0, as shown in Fig.~4(a),
$C_{\rm s}$ and $C_{\rm o}$ exhibit exactly the same behavior
with a four-site periodicity,
which is in agreement with the previous investigations
for the SU(4) exchange model.~\cite{Yamashita1998,Frischmuth1999}
For finite $J$, however,
$C_{\rm s}$ and $C_{\rm o}$ are not equivalent any longer
due to the breakdown of the SU(4) symmetry.
In Fig.~4(b),
it is found that $C_{\rm s}$ keeps the four-site periodicity,
while the amplitude of the oscillation becomes large.
On the other hand,
as shown in Fig.~4(c),
we observe a drastic change in the structure of $C_{\rm o}$.
Indeed, the four-site periodicity almost disappears,
and the correlation decays much faster for long distances
in comparison with that at $J$=0.

\begin{figure}[t]
\includegraphics[width=\linewidth]{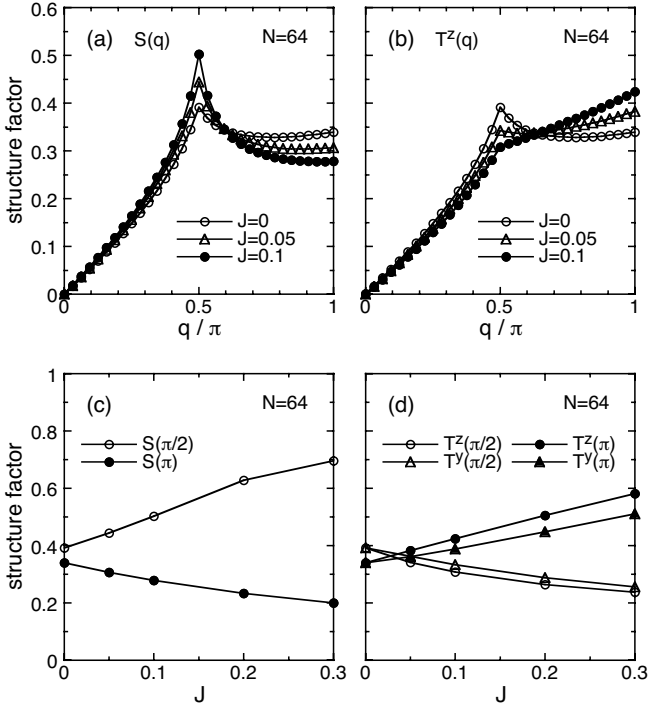}
\caption{
(a) Spin and (b) orbital structure factors at several values of $J$.
(c) Spin and (d) orbital structure factors at $q$=$\pi/2$ and $q$=$\pi$
as a function of $J$.
}
\end{figure}

In order to see clearly the changes in the spin and orbital structures,
it is useful to measure spin and orbital structure factors,
defined by
\begin{eqnarray}
 S(q)
 &=&
 \sum_{j,k}
 \langle S_j^z S_k^z \rangle
 e^{iq(j-k)}/N,
\\
 T^z(q)
 &=&
 \sum_{j,k}
 \langle T_j^z T_k^z \rangle
 e^{iq(j-k)}/N,
\\
 T^y(q)
 &=&
 \sum_{j,k}
 \langle T_j^y T_k^y \rangle
 e^{iq(j-k)}/N.
\end{eqnarray}
In Figs.~5(a) and 5(b),
we show $S(q)$ and $T^z(q)$ at several values of $J$, respectively.
At $J$=0,
$S(q)$ and $T^z(q)$ agree with each other
and have a peak at $q$=$\pi/2$ due to the clear four-site periodicity.
Even when $J$ is increased,
the peak position of $S(q)$ remains at $q$=$\pi/2$,
since $S(\pi/2)$ is enhanced and $S(\pi)$ is suppressed.
The $J$ dependence of $S(\pi/2)$ and $S(\pi)$ is plotted in Fig.~5(c).
On the other hand,
the peak position of $T^z(q)$ changes from $q$=$\pi/2$ to $q$=$\pi$,
since $T^z(\pi/2)$ is suppressed and $T^z(\pi)$ is enhanced,
in sharp contrast to the case of $S(q)$.
In fact, we find that $T^z(\pi)$ becomes larger than $T^z(\pi/2)$
at around $J$=0.05, as shown in Fig.~5(d).
It is interesting to note that
these characteristic changes of the spin and orbital structure factors
are naturally understood from the discussion for the four-site chain
in the previous subsection.
Namely, the spin-orbital configuration in Fig.~3(a) is stabilized
by the Hund's rule coupling at least for short distances,
leading to the enhancement of $S(\pi/2)$ and $T^z(\pi)$
in the whole system.

We note here that
$T^y(q)$ shows qualitatively the same behavior with $T^z(q)$,
but there appears a slight difference between them.
In Fig.~5(d),
$T^z(q)$ and $T^y(q)$ at $q$=$\pi/2$ and $q$=$\pi$
are plotted as a function of $J$.
With increasing $J$,
$T^y(\pi/2)$ is suppressed and $T^y(\pi)$ is enhanced
in the same way as $T^z(q)$,
while $T^z(\pi)$ is much enhanced than $T^y(\pi)$.
This anisotropic behavior is naturally explained
by the anisotropy in the orbital exchanges
$(J_{T}^{zx},J_{T}^{y},J_{ST}^{y})$
within a mean-field discussion.
For the $J_{ST}^{y}$ term
of the spin-orbital model (\ref{eq:H-so}), replace
${\bf S}_i \cdot {\bf S}_j$ by
$\langle {\bf S}_i \cdot {\bf S}_j \rangle$.
In the fully polarized FM state,
$\langle {\bf S}_i \cdot {\bf S}_j \rangle$
takes the maximum value 1/4 at all bonds,
and the anisotropy disappears,
since the relation
$J_{T}^{y}+J_{ST}^{y}/4=J_{T}^{zx}$
holds, as shown in Eq.~(\ref{eq:relation}).
However, in the spin-singlet state,
$\langle {\bf S}_i \cdot {\bf S}_j \rangle$
should take a smaller value than 1/4,
indicating that the orbital correlation in the $zx$ plane
becomes larger than that along the $y$ direction.

\begin{figure}[t]
\includegraphics[width=\linewidth]{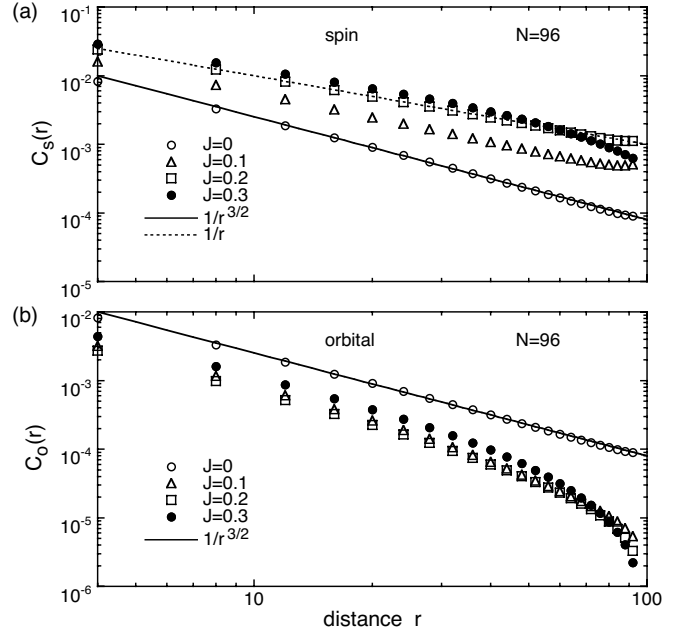}
\caption{
Log-log plot of (a) spin and (b) orbital correlation functions
at several values of $J$.
The results are shown for distances of four multiples.
}
\end{figure}

In Figs.~6(a) and 6(b), we show $C_{\rm s}$ and $C_{\rm o}$
in the log-log scale
to observe the decaying behavior.
At $J$=0,
$C_{\rm s}$ and $C_{\rm o}$ exhibit power-law decay
with the critical exponent 3/2,
consistent with the previous analytical works~\cite{Affleck1986}
and numerical investigations.~\cite{Yamashita1998,Frischmuth1999}
With increasing $J$,
as shown in Fig.~6(a),
$C_{\rm s}$ gradually increases
in accordance with the enhancement of $S(\pi/2)$,
while the slope seems to be gentle.
In particular,
we observe that $C_{\rm s}$ decays as $1/r$ at $J$=0.2,
as denoted by the dotted line in Fig.~6(a).
At $J$=0.3,
$C_{\rm s}$ comes to decay exponentially,
indicating the behavior of a quantum disordered spin state.
On the other hand,
as shown in Fig.~6(b),
$C_{\rm o}$ turns to show an exponential decay for smaller $J$
in comparison with $C_{\rm s}$,
suggesting that a quantum disordered orbital state is
stabilized by the Hund's rule coupling.

However,
it is important to remark that,
according to the generalized Kosterlitz-Thouless
transition,~\cite{Lee2004,Itoi1999}
the correlation length would diverge as
$\xi$$\sim$$\exp({\rm const.}/J)$ for small $J$.
Unfortunately, from numerical calculations,
it is quite difficult to distinguish whether
the correlation length is finite but very large
or certainly infinite.
In this sense,
our results indicate that the spin correlation develops
at least within the present system size,
and the variation of the power-law decay
could be a crossover phenomenon
from the SU(4) critical state to a quantum disordered spin state
due to the finite size effect.

\begin{figure}[t]
\includegraphics[width=\linewidth]{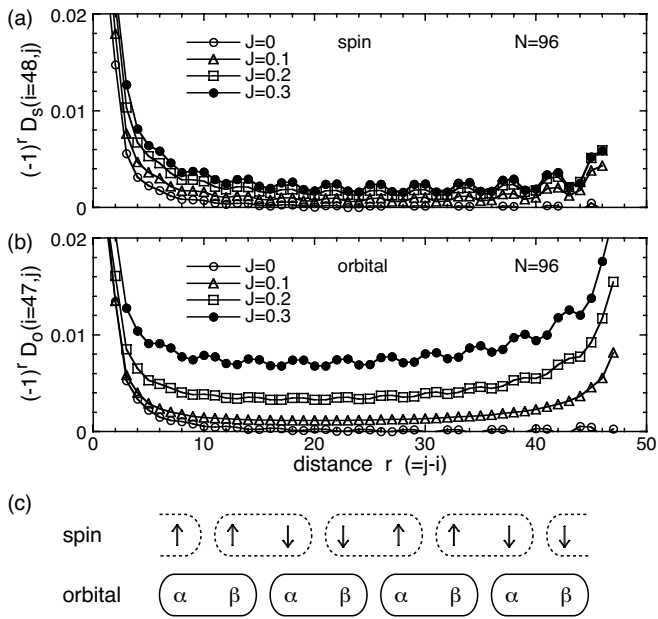}
\caption{
(a) Spin and (b) orbital dimer correlation functions,
measured from the center of the chain,
at several values of $J$.
(c) Spin-orbital configuration in the dimer state.
Dotted and solid oval enclosures represent
spin- and orbital-singlet pairs, respectively.
}
\end{figure}

Concerning the order parameter at finite $J$,
it has been proposed that
there occurs a dimer order with the alternating pattern of
spin- and orbital-singlet pairs,~\cite{Lee2004}
which is similar to the dimer state of
the SU(2)$_{\rm spin}$$\times$SU(2)$_{\rm orbital}$
exchange model.~\cite{Kolezhuk1998}
In order to clarify how spin and orbital dimerizations develop
due to the Hund's rule coupling,
we investigate spin and orbital dimer correlation functions,
defined by
\begin{eqnarray}
 D_{\rm s}(i,j)
 &=&
 \langle S_i^z S_{i+1}^z
       ( S_j^z S_{j+1}^z - S_{j+1}^z S_{j+2}^z ) \rangle,
\\
 D_{\rm o}(i,j)
 &=&
 \langle T_i^z T_{i+1}^z
       ( T_j^z T_{j+1}^z - T_{j+1}^z T_{j+2}^z ) \rangle.
\end{eqnarray}
In Figs.~7(a) and 7(b), $D_{\rm s}$ and $D_{\rm o}$
are shown, respectively, for several values of $J$.
Note that, taking into account the alternation of
the spin and orbital dimerizations,
we shift the starting point of $D_{\rm s}$ and $D_{\rm o}$ by one,
i.e., we set $i$=48 for $D_{\rm s}$ and
$i$=47 for $D_{\rm o}$ with 96 sites.
At $J$=0,
we find no indication of dimer long-range order,
since both $D_{\rm s}$ and $D_{\rm o}$ rapidly decay
as the distance becomes large.
With increasing $J$,
$D_{\rm s}$ and $D_{\rm o}$ show a tendency to grow in the system,
but we find a clear difference between them.
It is observed that
$D_{\rm s}$ remains small even when $J$ is increased,
while $D_{\rm o}$ shows a significant development,
indicating that the orbital dimerization is stabilized
by the Hund's rule coupling.

In Fig.~7(c), we schematically depict
the spin-orbital configuration in the dimer state.
Here, we note again that the Hund's rule coupling induces
FM and AFO interactions.
Indeed, with increasing $J$,
the ground state turns to be a fully polarized FM state
and the orbital state is described by a $T$=1/2 AFO critical state.
However, the ground state remains spin singlet for small $J$.
It is considered that
orbital-singlet dimers are formed to stabilize
a parallel spin configuration in each orbital-singlet dimer,
while the correlations between the dimers are suppressed.
Then, there occurs a staggered configuration of parallel spin pairs,
leading to the spin-singlet ground state.
Note that such a spin configuration is already seen
in the four-site problem, as shown in Fig.~3(a).

\begin{figure}[t]
\includegraphics[width=\linewidth]{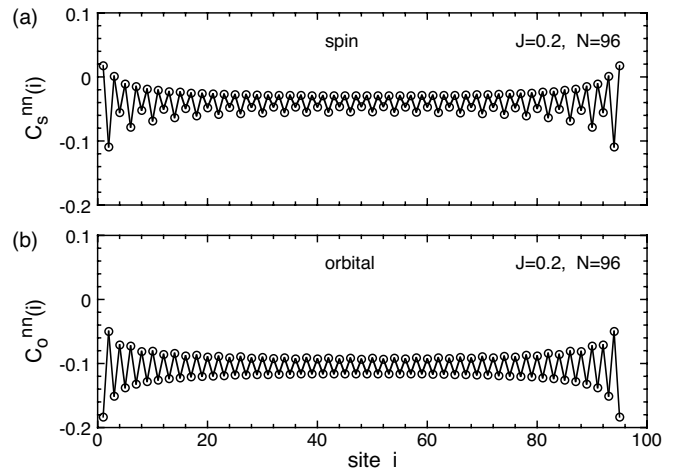}
\caption{
(a) Spin and (b) orbital correlation functions
between nearest-neighbor sites at $J$=0.2.
}
\end{figure}

We note that the dimer order is associated with
a spontaneously broken translational symmetry,
and the ground state has two-fold degeneracy
with a finite excitation gap.
The broken symmetry is discrete rather than continuous rotational symmetry,
so that a true long-range order can exist even in one dimension.
On the other hand,
in the present calculations,
one of the two possible dimerization patterns is favored
due to open boundaries.
In fact, we can see characteristic spatial patterns
in spin and orbital correlation functions
between nearest-neighbor sites,
\begin{eqnarray}
 C_{\rm s}^{\rm nn}(i)
 &=&
 \langle S_i^z S_{i+1}^z \rangle,
\\
 C_{\rm o}^{\rm nn}(i)
 &=&
 \langle T_i^z T_{i+1}^z \rangle.
\end{eqnarray}
As shown in Fig.~8,
we notice that $C_{\rm s}^{\rm nn}$ and $C_{\rm o}^{\rm nn}$
exhibit an oscillation in an anti-phase manner.
In particular, truly FM/AFO bonds appear around the edges.
As shown in Fig.~8(b),
orbital-singlet dimers are robustly formed at the edges
and the whole system is efficiently covered with
the orbital-singlet dimers,
so as to lower the energy making use of quantum orbital fluctuation.
Then, weakly formed spin-singlet dimers are arranged
alternately to the orbital-singlet dimers,
as shown in Fig.~7(c).

\begin{figure}[t]
\includegraphics[width=\linewidth]{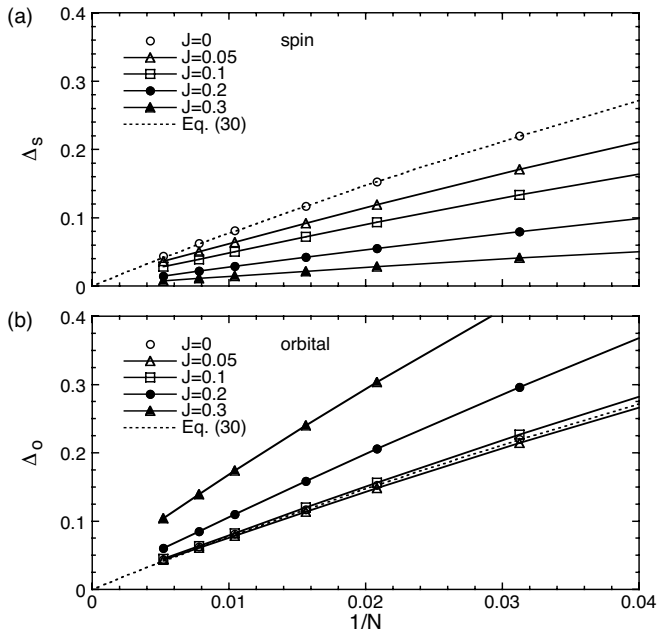}
\caption{
System size dependence of (a) spin and (b) orbital gaps
at several values of $J$.
Dotted curve denotes the analytical result at $J$=0,
given by Eq.~(\ref{eq:gap-exact}).
}
\end{figure}

Now we turn our attention to the low-lying excitation.
To clarify how the distinct behavior of spin and orbital states
is reflected in the spin and orbital excitations,
we investigate spin and orbital gaps,
\begin{eqnarray}
 \Delta_{\rm s} &=& E(1,0)-E(0,0),
\\
 \Delta_{\rm o} &=& E(0,1)-E(0,0),
\end{eqnarray}
where $E(S_{\rm tot}^z,T_{\rm tot}^y)$ denotes
the lowest energy in the subspace with
$S_{\rm tot}^z$ and $T_{\rm tot}^y$.
In Figs.~9(a) and 9(b), we show the system size dependence of
$\Delta_{\rm s}$ and $\Delta_{\rm o}$, respectively.
We should note that, at the SU(4) point $J$=0,
the elementary excitation is gapless,
while the finite size spectrum includes
a logarithmic correction,\cite{Itoi2000}
\begin{equation}
\label{eq:gap-exact}
 \Delta_{\rm s}=\Delta_{\rm o}=
 \frac{\pi^2}{N} \left( 1-\frac{1}{\ln N} \right).
\end{equation}
Indeed, it is observed that
$\Delta_{\rm s}$ and $\Delta_{\rm o}$ at $J$=0 are well reproduced by
Eq.~(\ref{eq:gap-exact}),
as denoted by the dotted curve.
With increasing $J$,
we find that $\Delta_{\rm s}$ becomes small for each system size
up to $N$=192, as shown in Fig.~9(a).
Even though the ground state is the gapped dimer state,
the spin gap does not show a remarkable development,
since the formation of the spin dimer state is very weak.
This result is consistent with the analytical work stating that
the spin gap in the thermodynamic limit is exponentially small as
$\Delta_{\rm s}$$\sim$$\exp(-{\rm const.}/J)$.~\cite{Lee2004}

On the other hand, as shown in Fig.~9(b),
we observe a quite different behavior of $\Delta_{\rm o}$.
At the beginning of the increase of $J$ ($J$$\alt$0.1),
$\Delta_{\rm o}$ keeps almost the same finite size spectrum.
For larger $J$,
$\Delta_{\rm o}$ shows a drastic increase for each system size,
and $\Delta_{\rm o}$ seems to converge to
a finite value in the thermodynamic limit.
We mention that such a development of the orbital gap
is accompanied by the stabilization of the orbital dimer state
due to the Hund's rule coupling.
Thus, the low-energy physics is described
by spin excitations below the orbital gap.
We conclude that
the Hund's rule coupling leads to the characteristic difference
between spin and orbital states,
and the magnetic property is represented by an effective spin system
on the background of the robust orbital dimer state.

\section{Summary}

In this paper,
we have studied the effect of the Hund's rule coupling
on the SU(4) spin-orbital exchange model.
The Hund's rule coupling introduces
isotropic FM and anisotropic AFO interactions
as well as the spin-orbital coupled term,
leading to the symmetry breakdown from SU(4) to
SU(2)$_{\rm spin}$$\times$U(1)$_{\rm orbital}$.
When the Hund's rule coupling is zero,
spin and orbital correlations coincide with each other
and have a peak at $q$=$\pi/2$
due to the SU(4) singlet ground state.
With increasing the Hund's rule coupling,
the peak position of orbital correlation is found to change
to $q$=$\pi$, while that of spin correlation remains at $q$=$\pi/2$,
due to the change of the relevant spin-orbital configuration.

On the other hand,
the Hund's rule coupling leads to the characteristic quantum behavior
in a different way between spin and orbital states.
In fact,
the orbital dimer state is stabilized by the Hund's rule coupling,
indicating the enhancement of quantum orbital fluctuation.
In accordance with the robust formation of the orbital dimer state,
we observe a development of the orbital gap,
while the spin gap is significantly suppressed
even in the gapped dimer state.
Thus, the low-energy physics is described by
a nearly critical spin state with the small spin gap
on the background of the robust orbital dimer state.
We believe that the present study is useful to explore novel quantum states
in actual low-dimensional materials with active orbital degree of freedom.


\begin{acknowledgments}
The author is grateful to T. Hotta and K. Ueda
for useful discussions and comments.
He also thanks K. Kubo for comments.
He is supported by the HOUGA Research Resources of
Japan Atomic Energy Agency
and the Grant-in-Aid for Scientific Research
in Priority Area ``Skutterudites''
from the Ministry of Education, Culture, Sports, Science, and
Technology of Japan.
\end{acknowledgments}


\end{document}